\documentclass[journal]{IEEEtran}

\usepackage{gensymb}

\ifCLASSINFOpdf
   \usepackage[pdftex]{graphicx}
\else
\fi
  \usepackage[caption=false,font=normalsize,labelfont=sf,textfont=sf]{subfig}
\begin{document}
\title{Tunable MEMS VCSEL on Silicon substrate}

%\author{Hitesh~Kumar~Sahoo,~\IEEEmembership{Member,~IEEE,}
\author{Hitesh~Kumar~Sahoo, Thor~Ansb{\ae}k, Luisa~Ottaviano, Elizaveta~Semenova,  Fyodor Zubov, Ole~Hansen,  and~Kresten~Yvind% <-this % stops a space
\thanks{H.K.Sahoo is with DTU Fotonik, Technical University of Denmark (DTU), Kongens Lyngby DK-2800, e-mail: hikus@fotonik.dtu.dk.}% <-this % stops a space
\thanks{T. Ansb{\ae}k and L. Ottaviano were with DTU Fotonik, DTU and are now with OCTLight ApS and Alight Aps respectively.}% <-this % stops a space
\thanks{E. Semenova and O. Hansen are  with DTU Fotonik and DTU Nanotech, respectively, at DTU.}
\thanks{F. Zubov is with Saint Petersburg Academic University,  Russia.}
\thanks{K. Yvind is  with DTU Fotonik, DTU, e-mail: kryv@fotonik.dtu.dk.}
\thanks{Manuscript received ??; revised ??.}}


\markboth{Journal of \LaTeX\ Class Files,~Vol.~??, No.~+, ??~2019}%
{Shell \MakeLowercase{\textit{et al.}}: Bare Demo of IEEEtran.cls for IEEE Journals}



\maketitle

\begin{abstract}
%We present the design, fabrication and characterization of a MEMS VCSEL which utilized a silicon on insulator wafer for the microelectromechanical system and encapsulates the MEMS by direct InP wafer bonding in order to improve the protection and control of the tuning element. This can enable more robust fabrication, a larger free spectral range and bi-directional tuning of the MEMS element. The proposed device uses a high contrast grating mirror on a MEMS stage as the bottom mirror, wafer bondind InP with quantum wells for amplification and a deposited dielectric DBR. A tuning range of 40 nm and a mechanical resonance frequency of $>$2MHz is demonstrated.
We present design, fabrication and characterization of a MEMS VCSEL which utilizes a silicon-on-insulator wafer for the microelectromechanical system and encapsulates the MEMS by direct InP wafer bonding, which improves the protection and control of the tuning element. This procedure enables a more robust fabrication, a larger free spectral range and facilitates bidirectional tuning of the MEMS element. The MEMS VCSEL device uses a high contrast grating mirror on a MEMS stage as the bottom mirror, a wafer bonded InP with quantum wells for amplification and a deposited dielectric DBR as the top mirror. A 40 nm tuning range and a mechanical resonance frequency in excess of 2 MHz are demonstrated.
\end{abstract}


% \fi
\IEEEpeerreviewmaketitle

\section{Introduction}

\IEEEPARstart{O}{ptical} Coherence Tomography (OCT) has grown rapidly %in a span of 26 years 
since its introduction in 1991 \cite{drexler2014optical,huang1991optical}. Imaging resolution and speed have increased by 10 and 1 000 000 times, respectively \cite{drexler2014optical}. The application area which initially was eye imaging \cite{huang1991optical} has now expanded to cardiology \cite{bezerra2009intracoronary,suter2011intravascular}, dermatology \cite{welzel2001optical}, and urology \cite{tearney1997optical} etc. Integration of OCT with endoscopes \cite{fujimoto2003optical,liu2004rapid} and surgical probes \cite{boppart1997forward} for providing physicians with real-time data has also been shown. OCT has also been coupled with other imaging technologies such as multiphoton tomography \cite{konig2009clinical}, non-linear microscopy \cite{vinegoni2004nonlinear}, and photoacoustic imaging \cite{zhang2011multimodal} to extract more enriched information. In order to continue this trend and extent to more application domains, OCT systems need to be simplified. The ultimate goal should be to integrate OCT on a chip, in a compact form factor. The light source is the most important component of the OCT system, and the MEMS vertical-cavity surface-emitting laser (VCSEL) based swept sources (SS) \cite{ansbaek2013resonant,jayaraman2012high} are among the most promising light source technologies for future integrated OCT systems.

Most existing MEMS VCSELs, e.g., \cite{ansbaek2012vertical,ansbaek20131060,jayaraman2012rapidly,rao2013long} primarily based on the III-V material system, share a common design principle. The bottom mirror is grown along with the active material and the actuated part is the top mirror. As a result, they have an open MEMS design and need careful handling until hermetically sealed using an external packaging. Changes in the humidity and temperature may result in water condensation on the device which damages the MEMS element. A sealed MEMS element would make a more robust device. Moreover, the MEMS element can be actuated in one direction only since the electrostatic force is an attractive force. Also, the MEMS element can only be actuated one third of the total gap distance to prevent electrostatic snap-in. Consequently the gap must be increased in order to increase the tuning range, however, the increased gap also results in a longer laser cavity and thereby a smaller free space range (FSR). So, there is always a trade-off between a short laser cavity and a wide mirror actuation range. This also sets a limit to mode-hop-free tunable wavelength bandwidth.  Thus these issues need to be addressed in order to design a more robust and better tunable MEMS VCSEL.

In the following we propose a MEMS VCSEL design based on a Silicon substrate which incorporates sealed MEMS actuation and the possibility to decouple the actuation gap from the laser cavity. The bottom mirror of the VCSEL is the actuated component defined on a Silicon-on-insulator (SOI) substrate, which is then bonded to a III-V wafer with active material followed by a deposition of the top mirror. The design of the MEMS VCSEL is discussed in Sec. \ref{sec:Design}. The fabrication process flow is presented in Sec. \ref{sec:fab} followed by characterization of an optically pumped 1550 nm MEMS VCSEL in Sec. \ref{sec:char}.

\section{Design of MEMS VCSEL on Silicon substrate\label{sec:Design} }
\begin{figure}[!t]
\centering
\includegraphics[width=3.5in]{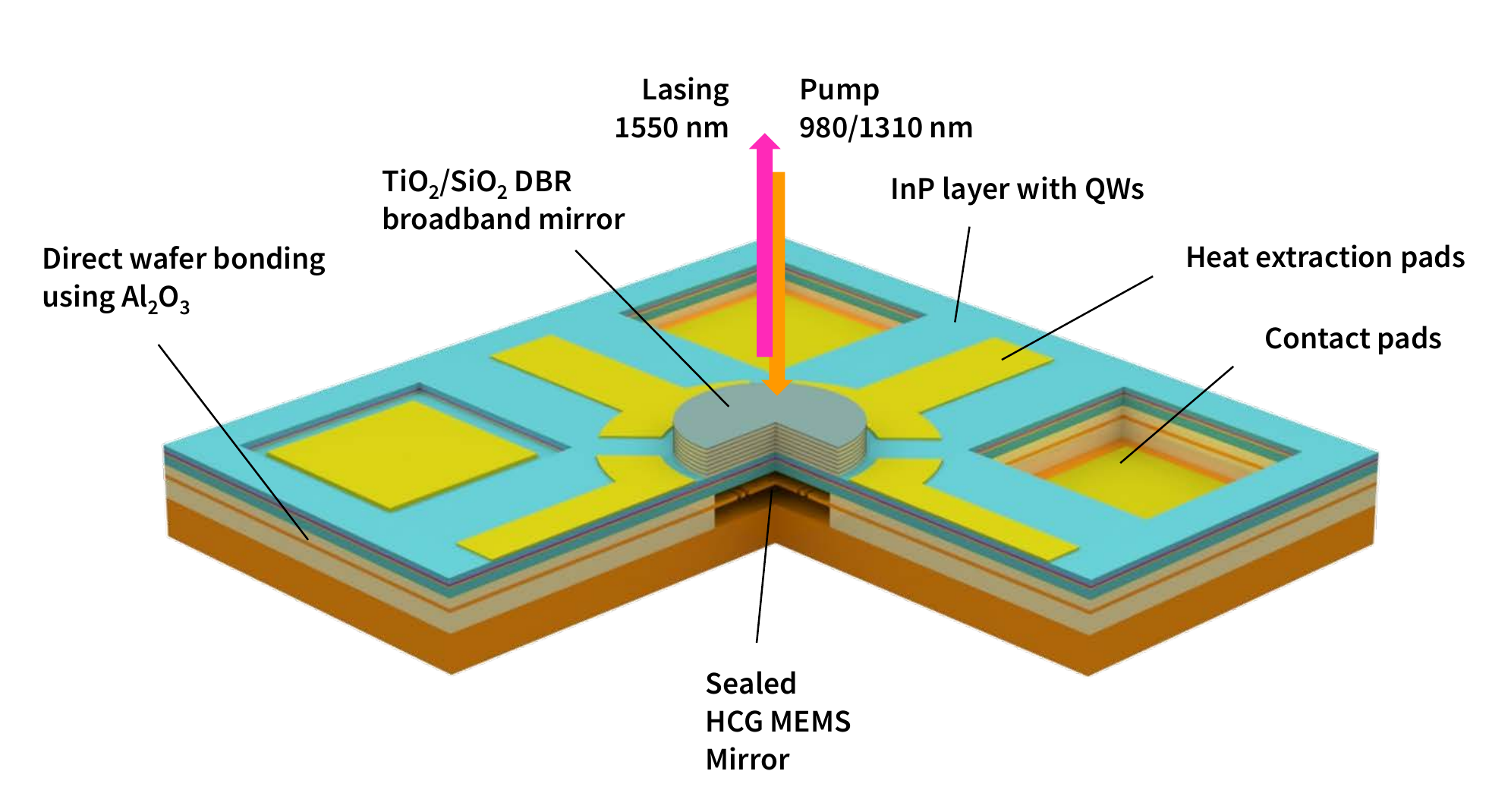}
%where an .eps filename suffix will be assumed under latex, 
% and a .pdf suffix will be assumed for pdflatex; or what has been declared
% via \DeclareGraphicsExtensions.
\caption{Illustration of a MEMS VCSEL on a Silicon substrate. The cut-out section shows the sealed MEMS cavity.}
\label{fig_crosssection}
\end{figure}

%The schematic of a MEMS VCSEL on a Silicon substrate is shown in Fig. \ref{fig_crosssection}. A high contrast grating (HCG) mirror \cite{chang2010high} is used as the bottom mirror. It sits on a MEMS frame which is defined on the Silicon "device" layer of an SOI substrate. The active material and the top mirror, dielectric distributed Bragg reflector (DBR) are static and seal the MEMS cavity. The DBR and HCG mirror are now not dependent on a challenging epitaxial growth. On the contrary both the mirrors can be defined using standard complementary-metal-oxide-semiconductor (CMOS) processing steps. Last but not the least, the device is now based on a Silicon substrate which makes (automated) handling the wafer easier and clears the path for further integration.
Fig.~\ref{fig_crosssection} shows a schematic of the MEMS VCSEL on a Silicon substrate. A high contrast grating (HCG) mirror \cite{chang2010high} is used as the bottom mirror, which is fixed on a MEMS frame defined in the Silicon "device" layer of an SOI substrate. The active material and the top mirror, a dielectric distributed Bragg reflector (DBR), are static and seal the MEMS cavity. Thus, the DBR and HCG mirror are no longer dependent on a challenging epitaxial growth; on the contrary both mirrors can be defined using standard complementary-metal-oxide-semiconductor (CMOS) processing steps. Last but not the least, the device is now based on a Silicon substrate which makes (automated) handling of the wafer easier and that paves the path for further integration.

%There are many benefits to having a sealed cavity. There is a better control on the MEMS dynamics. Partial vacuum or gases can be introduced into the cavity to have a control on the MEMS damping. In addition, the HCG on the MEMS can now be actuated both from the top and the bottom, essentially increasing the tuning bandwidth. There is also the option to only actuate using the bottom electrode, thus realizing a pull away design. This decouples the relation between the tuning gap and the laser cavity as the HCG bottom mirror can be placed close to the active layer resulting in a wide FSR.  
A sealed cavity provides multiple benefits, such as improved control of the MEMS dynamics, since a partial vacuum or controlled gas pressure in the cavity can be used to adjust the MEMS damping. In addition, the HCG on the MEMS can now be actuated both from the top and the bottom electrodes, essentially increasing the tuning bandwidth. Actuation using only the bottom electrode is also an option, which realizes a pull-away design. This option decouples the relation between the tuning gap and the laser cavity since the HCG bottom mirror can be placed close to the active layer resulting in a wide FSR.  

\begin{figure}[!t]
\centering
\subfloat[]{\includegraphics[width=3in]{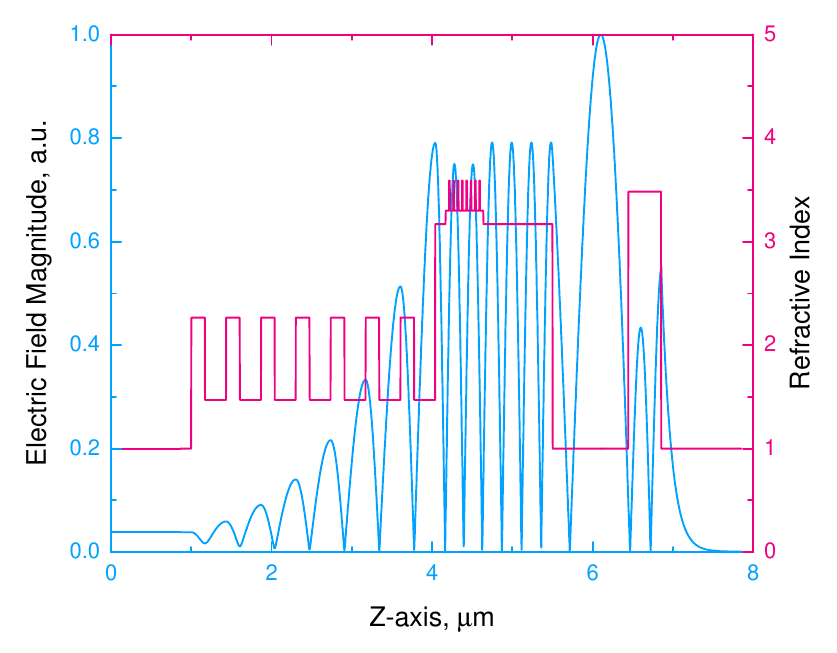}}%
\hfil
\subfloat[]{\includegraphics[width=3 in]{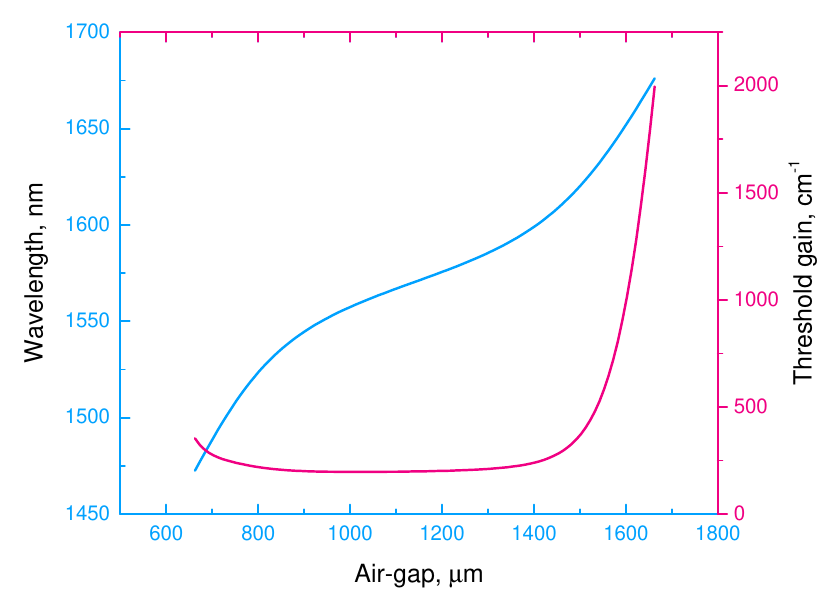}}%
\caption{(a) Electric field plotted on top of the device structure defined by
the refractive index of the layers. (b) The lasing wavelength and threshold gain plotted against the corresponding airgap (tunable component of the cavity length). }
\label{fig_sim}
\end{figure}

%The device was modeled using a modal method\cite{lavrinenko2014numerical} using CAMFR\cite{bienstman2001rigorous,bienstman2001optical}. Fig. \ref{fig_sim} (a) shows the refractive index profile of the defined structure overlapped on the resulting electric field. The alternating layers of SiO$_{2}$ and TiO$_{2}$ represent the top mirror, followed by a layer of active material, tunable airgap and HCG mirror. The active material was defined by 8 quantum wells (QWs) (InGaAsP/InGaAlAs) in InP.  The air-gap was varied to simulate the tuning effect. For each air-gap, the corresponding cavity mode and threshold gain were determined. The lasing mode was obtained by satisfying the condition of unity round trip gain for the cavity and the threshold gain was equal. Multiple QWs are used to minimize the impact of the moving antinotes of the field when the device is tuned so a the low threshold can be achieved for the desired range. A semiconductor coupled design was chosen for the device since the attempt to incorporate an AR coating in the bonding interface failed either due to roughness or stress in the coating. An air dominant design \cite{cook2019resonant} could also be done and since the HCG mirror is not the output mirror this could be done with minimum penalty from the needed high reflectivity of the MEMS mirror. 
The device was modeled using a modal method \cite{lavrinenko2014numerical} using CAMFR \cite{bienstman2001rigorous,bienstman2001optical}. Fig.~\ref{fig_sim}~(a) shows the refractive index profile of the structure overlaid on the resulting electric field. From the left the alternating layers of SiO$_{2}$ and TiO$_{2}$ represent the top mirror, followed by a layer of active material, the tunable airgap and the HCG mirror. The active material was defined by 8 quantum wells (QWs) (InGaAsP/InGaAlAs) in InP.  The air-gap thickness was varied to simulate the effect of tuning. For each air-gap thickness, the corresponding cavity mode and threshold gain were determined as shown in Fig.~\ref{fig_sim}~(b). The lasing mode was obtained by satisfying the condition of unity round trip gain for the cavity and the threshold gain was equal to minimum gain required to overcome losses of the system. Multiple QWs were used to minimize the impact of the moving anti-nodes of the field when the device is tuned thus a low threshold can be achieved for the desired wavelength range. A semiconductor coupled design was chosen for the device since the attempt to incorporate an AR coating at the bonding interface failed either due to roughness or stress in the coating. An air dominant design \cite{cook2019resonant} could also be done and since the HCG mirror is not the output mirror this could be done with minimum penalty from the needed high reflectivity of the MEMS mirror.

\section{1550 nm MEMS VCSEL Fabrication \label{sec:fab} }
\begin{figure}[!t]
\centering
\includegraphics[width=2.5in]{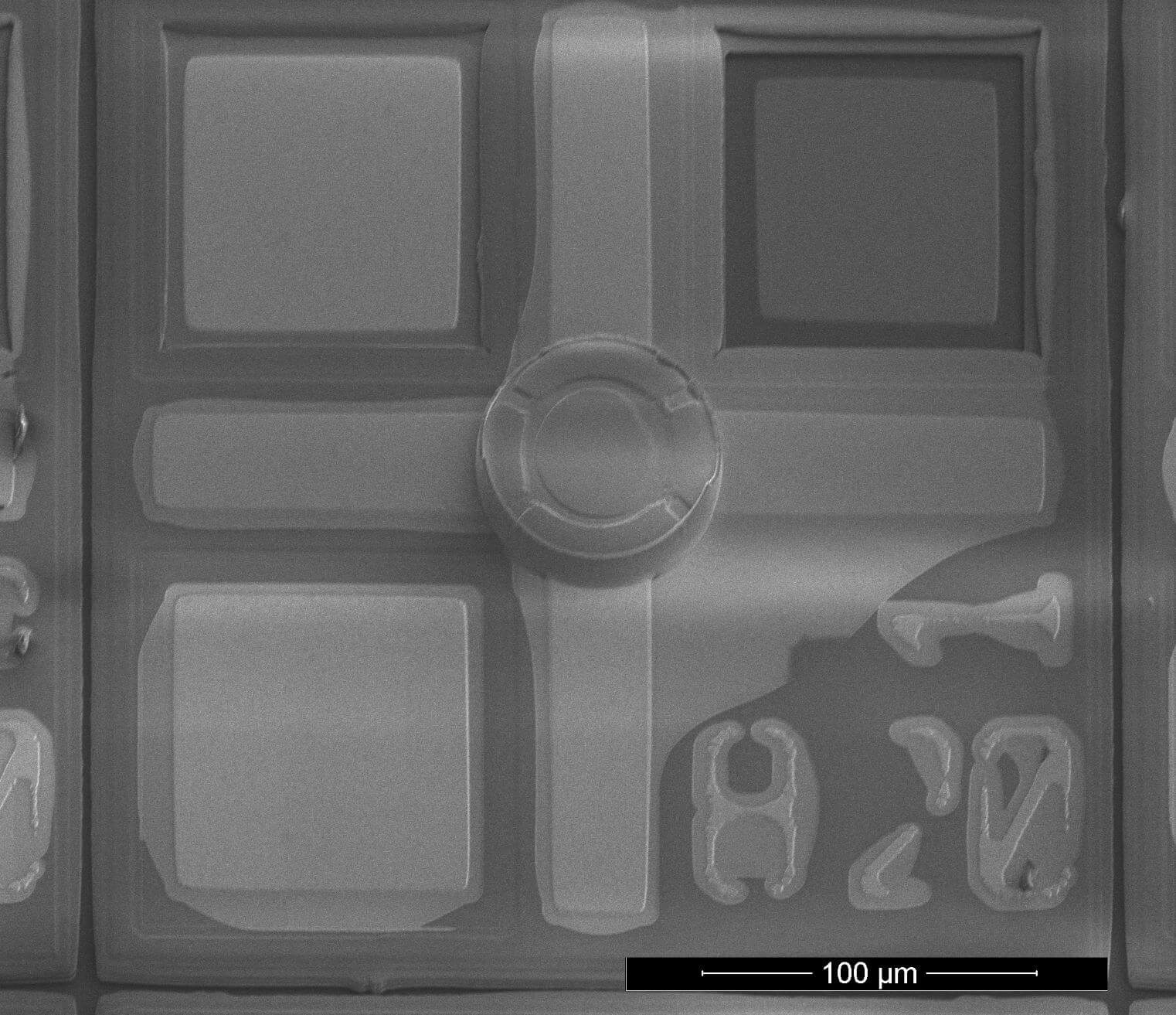}
\caption{SEM image of a fabricated MEMS VCSEL. The dark region around the metal pads is a thin layer of SiO$_{2}$ mask. The odd shape is due to an improper adhesion with the resist mask used for metal deposition.}
\label{fig_sem}
\end{figure}

Fig.~\ref{fig_fabfull} shows the fabrication process flow for the MEMS VCSEL. A SOI wafer with a 400$\pm$5 nm p-doped Silicon “device” layer (10 nm less than the optimal value for the HCG) on a 1~$\mu$m thick buried oxide was used as the substrate. Electron beam lithography was used to define the HCG and the MEMS frame around it. For a reflectivity around 99.9\% at 1550 nm, the HCG was defined with a periodicity of 650 nm and duty cycle of 0.63$\pm$0.08 to account for process variations.  The pattern was transferred to the MEMS Silicon layer using reactive ion etching (RIE). A blanket layer of SiO$_{2}$  was then deposited. The thickness of this layer equals the air-gap between the active layer and HCG mirror. SiO$_{2}$ was chosen since it is a standard CMOS material and easy to work with. This layer is also the interface layer for bonding with the active material, thus the surface roughness of the SiO$_{2}$ layer must be below 0.5 nm for successful direct bonding. However, SiO$_{2}$  deposited by plasma enhanced chemical vapor deposition (PECVD) has a surface roughness in excess of 3 nm, thus a two-step deposition process was used to achieve the required surface roughness. Following PECVD deposition of SiO$_{2}$, a layer of phosphorous boron doped silicate glass (PBSG) was deposited and then the stack was annealed at 1000 \degree C for re-flow of the glass. The re-flow of the PBSG resulted in a surface roughness of approximately 0.2 nm. The reason for using two types of SiO$_{2}$ was the relatively low wet etch rate of PBSG in buffered HF (bHF), which becomes critical in the subsequent etch step. A dry etch step followed by a wet etch using bHF was used to release the MEMS structure. The wet etch avoided ion damage on the HCG MEMS. Using only PBSG would increase the etch time in bHF which would degrade the resist mask. The wet etch in bHF is continued to etch also the buried oxide below the HCG MEMS. Critical point drying (CPD) step was used in order to successfully release the MEMS. The processed wafer was then direct bonded to an InP wafer with the active material, using Al$_{2}$O$_{3}$ as an intermediate layer \cite{sahoo2018low}.  The wafers were annealed at 300 \degree C to consolidate the wafer bonding. The InP substrate on the bonded wafers was then etched away leaving a thin layer of InP with the active region. Contact-windows were etched to all the contact layers – the n-doped layer in InP, the MEMS Silicon layer and the Silicon substrate. A combination of dry and wet etch steps were used in all three contact-window etches to avoid ion damage to the contact surface. A metal stack (25 nm  Ti, 75 nm Pt, 300 nm Au) was then deposited for establishing contacts using a lift-off process. In addition to the metal contacts, metal strips were added on top of the chip to improve with heat extraction from the lasing region (Visible in Fig.~\ref{fig_sem}). Finally, 7 pairs of TiO$_{2}$ (n = 2.31 at 1550 nm) and SiO$_{2}$ (n = 1.44 at 1550 nm) layers were deposited using sputter deposition to define the top DBR mirror. The individual layer thicknesses were optimized to reduce the reflectivity near the pump wavelengths while retaining high reflectivity around 1550 nm. A scanning electron microscope (SEM) image of a completed MEMS VCSEL is shown in Fig.~\ref{fig_sem}.

%The design of the MEMS VCSEL brings in flexibility into the fabrication process flow. All critical steps such as electron beam lithography, growth of active material and bonding are performed early in the process flow followed by the non-critical steps such as contact definition and DBR deposition. The bottom mirror (HCG on MEMS frame) and the top mirror (DBR) are defined using standard Si processing. 
The design of the MEMS VCSEL adds flexibility to the fabrication process flow. All critical steps such as electron beam lithography, growth of active material and bonding are performed early in the process flow followed by the non-critical steps such as contact definition and DBR deposition. The bottom mirror (HCG on MEMS frame) and the top mirror (DBR) are defined using standard Si processing.

\iftrue
\begin{figure}[!t]
\centering
\subfloat[]{\includegraphics[width=0.45\columnwidth]{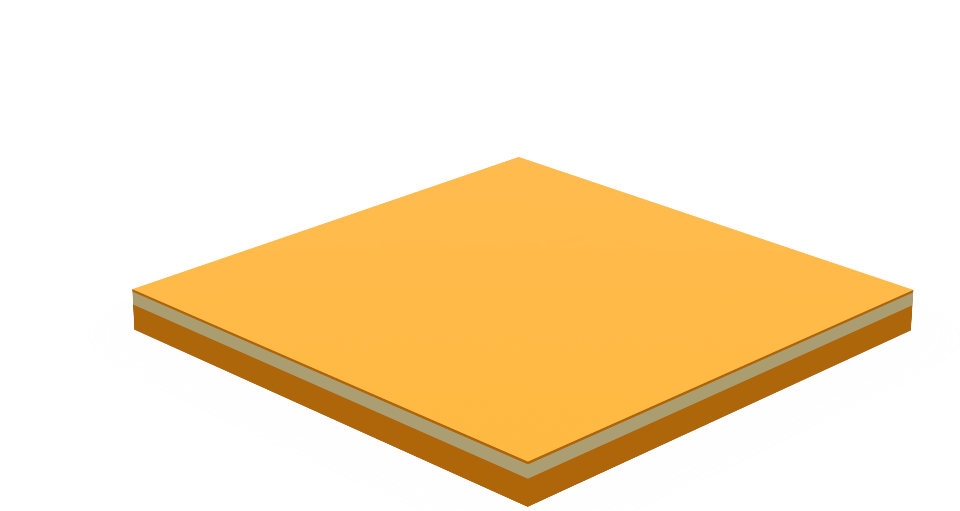}%
\label{SOI}}
\hfil
\subfloat[]{\includegraphics[width=0.45\columnwidth]{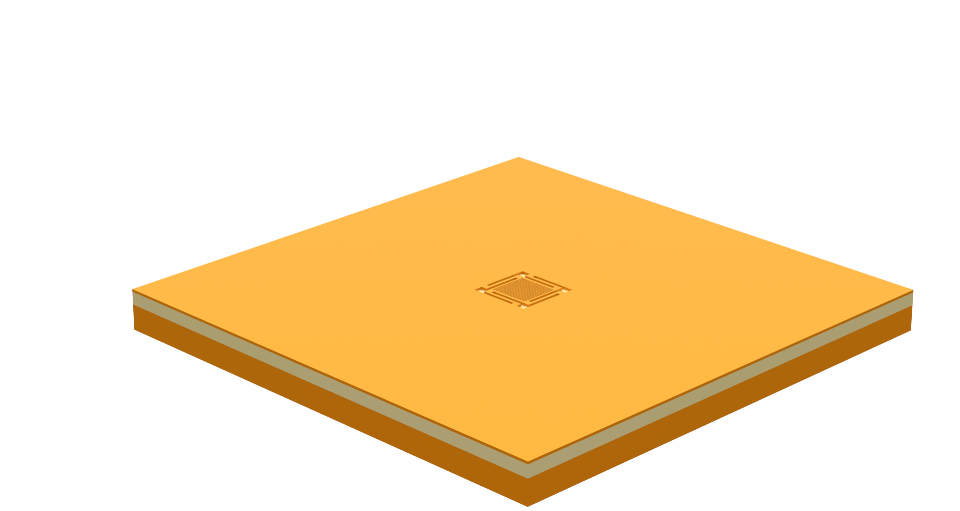}%
\label{ebeam_mems}}\\
\subfloat[]{\includegraphics[width=0.45\columnwidth]{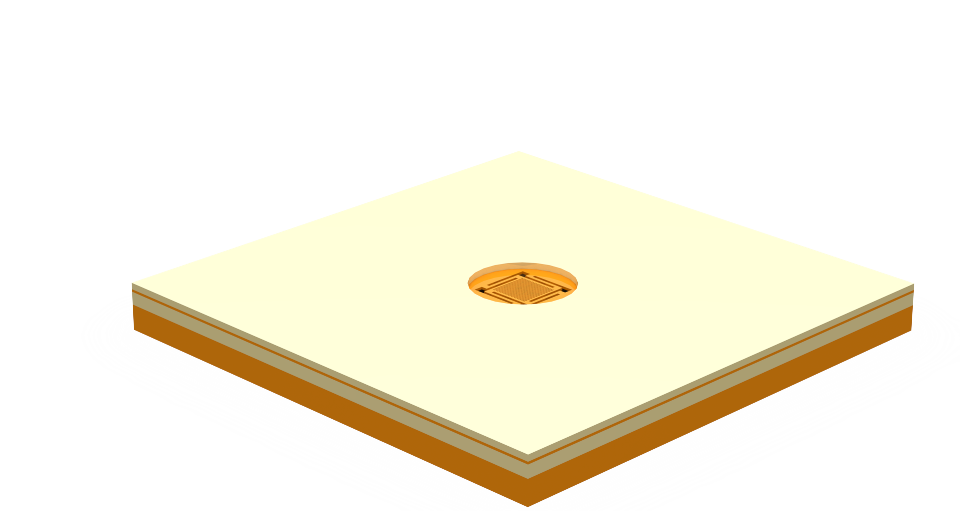}%
\label{mems_released}}
\hfil
\subfloat[]{\includegraphics[width=0.45\columnwidth]{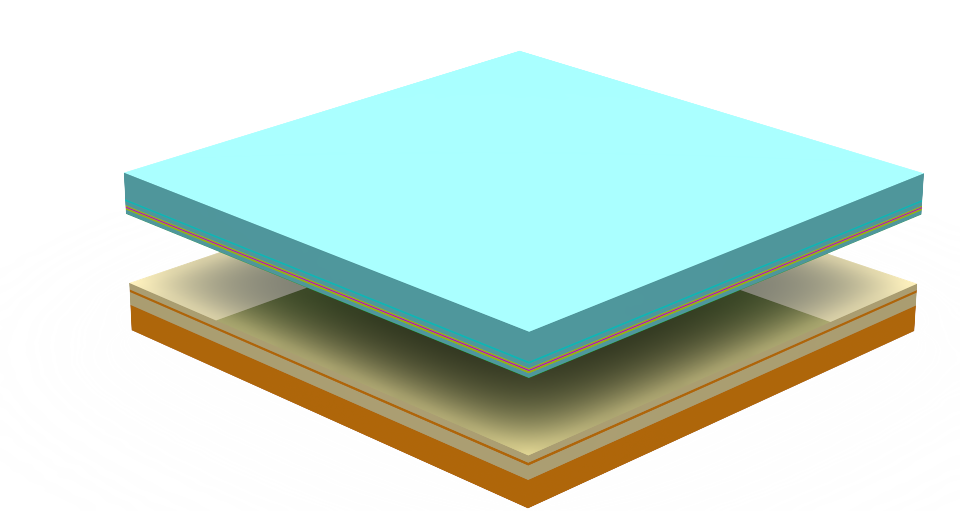}%
\label{directbonding}}\\
\subfloat[]{\includegraphics[width=0.45\columnwidth]{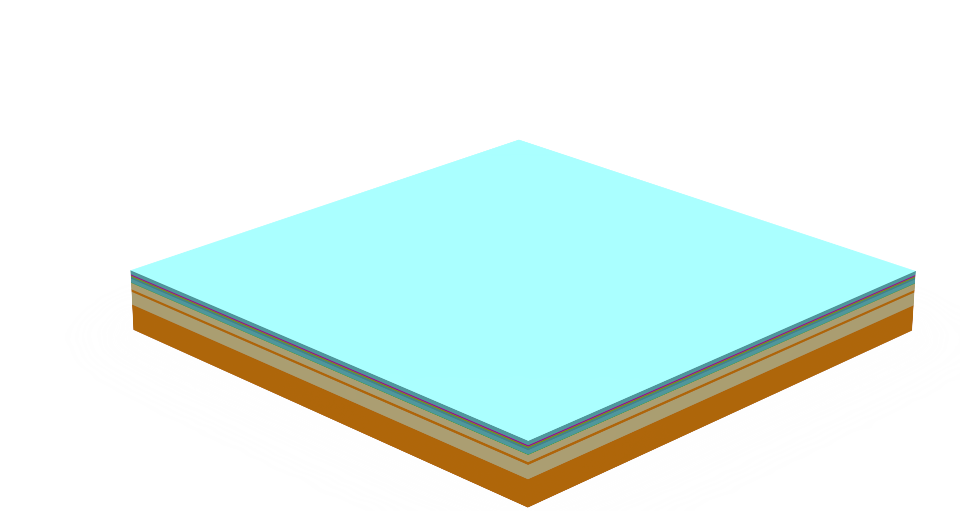}%
\label{substrate_removal}}
\hfil
\subfloat[]{\includegraphics[width=0.45\columnwidth]{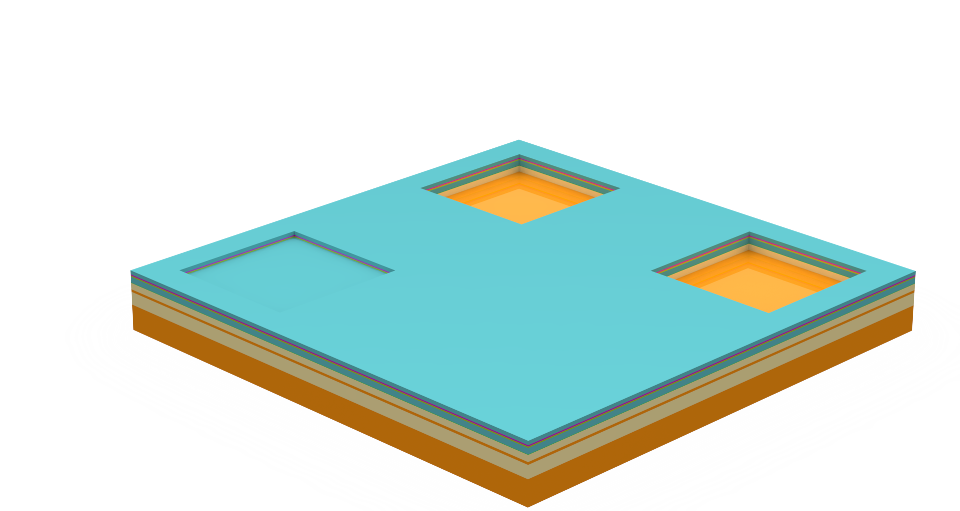}%
\label{etch_multi}}\\
\subfloat[]{\includegraphics[width=0.45\columnwidth]{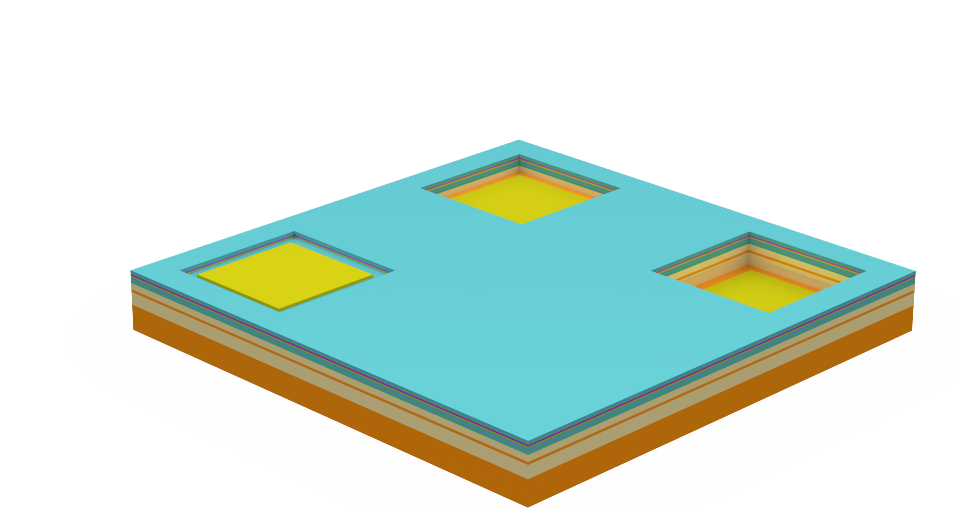}%
\label{Au_deposit}}
\hfil
\subfloat[]{\includegraphics[width=0.45\columnwidth]{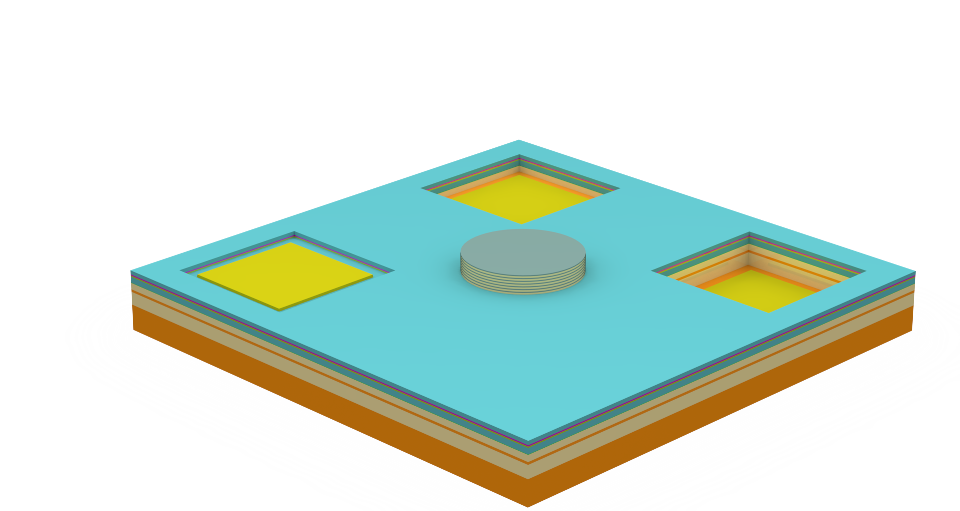}%
\label{DBR}}
\caption{Fabrication process flow of a 1550 nm MEMS VCSEL on a Silicon substrate. (a) A SOI wafer with a silicon device layer thickness of 400 nm was used as  the substrate. (b) The bottom mirror for the VCSEL, the HCG was defined in the silicon layer using e-beam lithography and RIE. (c) A blanket layer of 930 nm SiO$_{2}$  was then deposited followed by etch steps to open up on top of the patterned area and release the MEMS using CPD. (d) The processed wafer was directly bonded using Al$_2$O$_3$ as an intermediate layer to InP wafer with the QW structures. (e) The InP substrate was then etched away. (f) Openings were etched to all the contact layers - n-doped layer in InP, MEMS Silicon layer and the Silicon substrate. (g) Metal stack was deposited for establishing contacts. (h) 7 pairs of TiO$_{2}$ and SiO$_{2}$ layers were deposited to define the top DBR mirror.}
\label{fig_fabfull}
\end{figure}
\fi

\section{Device Characterization and Discussion \label{sec:char}}
\begin{figure}[!t]
\centering
\includegraphics[width=3in]{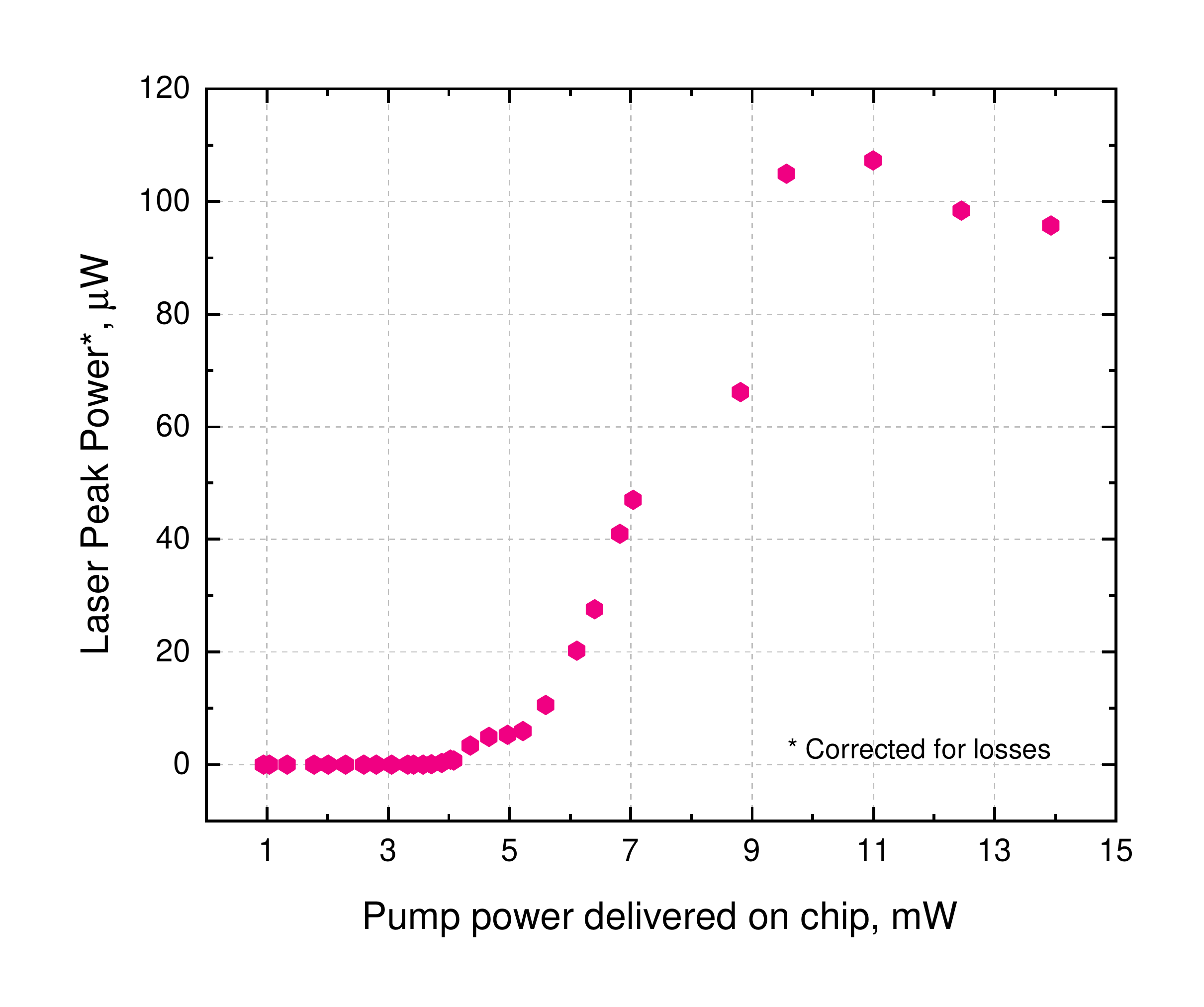}
\caption{Laser peak power plotted against the pump power. The collected laser peak power is corrected for the losses in the system.}
\label{fig_thres}
\end{figure}

%The fabricated device was optically pumped using a continuous wave 976 nm fiber Bragg grating laser diode. The pump was also used to define the aperture for lasing. A part of the incident pump energy is used for the generation of photons and the rest is released as heat. The heat increases refractive index at the incident spot and this defines the lasing mode. A microscope setup was used to focus the pump beam. Fig. \ref{fig_thres} shows threshold plot for a VCSEL at room temperature.  The threshold input power (delivered on the chip) is $\approx$5.5 mW which is translated to around 1.4 mW of absorbed power (assuming the active material has an absorption  co-efficient of 10$^{4}$ cm$^{-1}$), which is equivalent to an electrical injection current of 1 mA.  The maximum power output achieved for the device is 0.1 mW. There are small variations in Fig. \ref{fig_thres} due to undesired change in focus of the microscope.
The fabricated device was optically pumped using a continuous wave 976 nm fiber Bragg grating laser diode. The pump was also used to define the aperture for lasing; part of the incident pump energy is used for the generation of photons and the remaining is released as heat. The heat increases refractive index at the incident spot and this defines the lasing mode. A microscope setup was used to focus the pump beam. Fig. \ref{fig_thres} shows the threshold plot for a VCSEL at room temperature.  The threshold input power (delivered on the chip) is $\approx$5.5~mW which translates to around 1.4~mW of absorbed power (assuming the active material has an absorption coefficient of 10$^{4}$~cm$^{-1}$), which is equivalent to an electrical injection current of 1 mA.  The maximum power output achieved for the device is 0.1~mW. The small variations seen in Fig.~\ref{fig_thres} are due to undesired changes in focus of the microscope.

\begin{figure}[!t]
\centering
\includegraphics[width=3in]{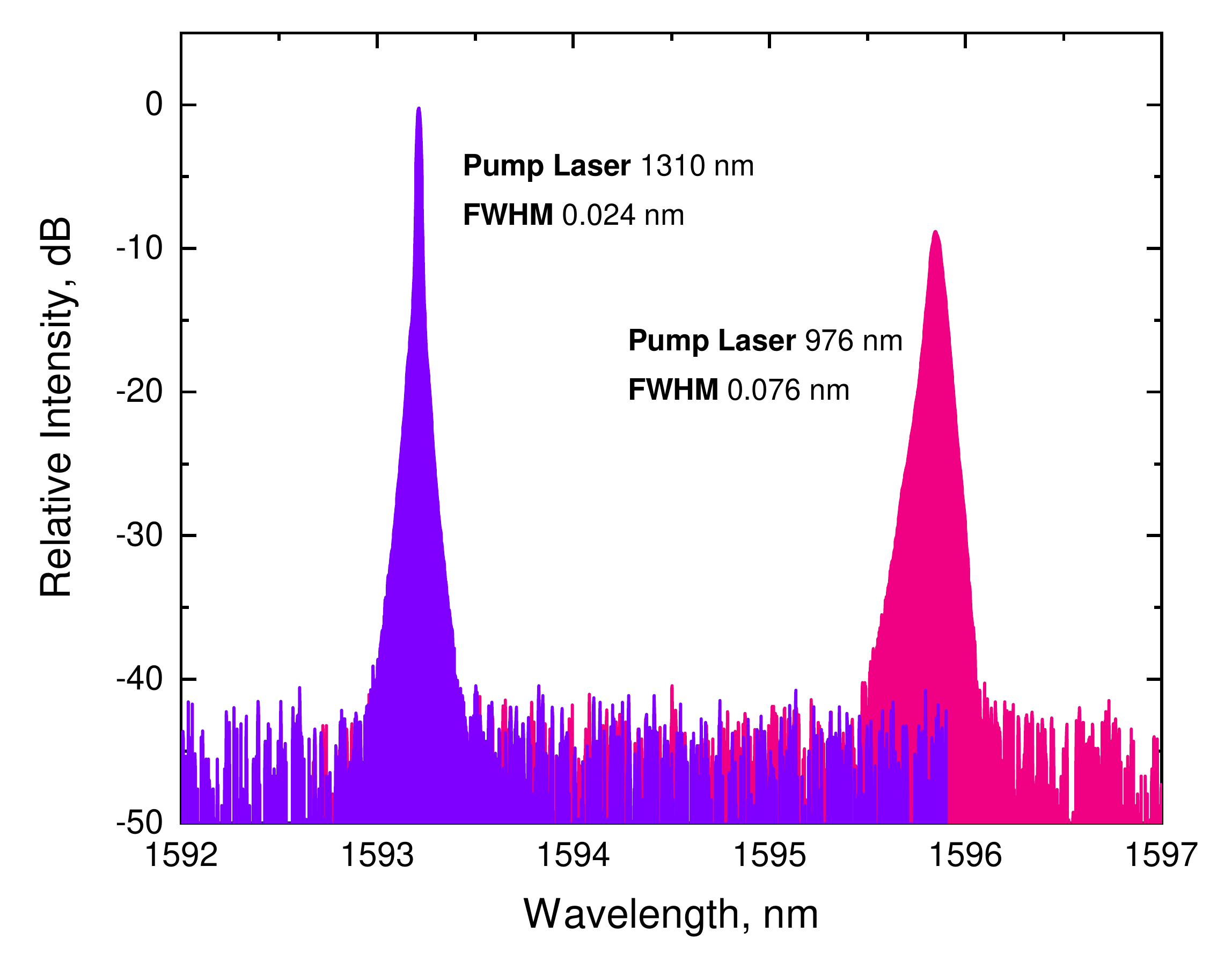}
\caption{Optical spectra obtained using 976 nm and 1310 nm pump lasers. The FWHM measured on the emission using the 1310 nm pump laser is smaller than that obtained using a 976 nm pump laser.  }
\label{fig_dual}
\end{figure}

%Fig. \ref{fig_spec}  shows the optical spectrum measured on the fabricated sample. It shows a single mode emission with peak at 1564 nm and a side mode suppression ratio of more than 55 dB. The use of the HCG as the bottom mirror also helps with the suppression of higher order modes.\cite{chung2008subwavelength} On the other hand, a HCG being defined in Silicon (which absorbs 976 nm) broadened the line-width of the emission due to heating as also reported by Rao et al.\cite{rao2013long} Fig. \ref{fig_heat} shows the optical spectrum obtained for 980 nm and 1310 nm pumping. (Note Fig. \ref{fig_spec}  and  Fig. \ref{fig_heat}  were performed on different devices.) The lasing peak corresponding to 980nm pump is red shifted due to heating. The measured full width half maximum (FWHM) was much smaller for a 1310 nm pumping. However, for the QWs used in the device, a 976 nm laser was more effective than 1310 nm laser as the former is absorbed both in the QWs and its barriers whereas the later is only absorbed in the available states within the QWs. So, a 980 nm laser was used for further characterization.
Fig.~\ref{fig_spec} shows the optical spectrum measured on the fabricated sample. The spectrum shows single mode emission with a peak at 1564 nm and a side mode suppression ratio of more than 55 dB. The use of the HCG bottom mirror also helps with suppression of higher order modes \cite{chung2008subwavelength}. On the other hand, a HCG defined in Silicon (which absorbs 976 nm) broadens the line-width of the emission due to heating as also reported by Rao et al.~\cite{rao2013long}. Fig.~\ref{fig_dual} shows the optical spectra obtained for 980 nm and 1310 nm pumping (Note, measurements in Figs.~\ref{fig_spec} and \ref{fig_dual} were performed on different devices). The lasing peak corresponding to 980 nm pumping is red shifted due to heating. The measured full width half maximum (FWHM) was much smaller using 1310 nm pumping. However, for the QWs used in the device, a 976 nm pump is more effective than a 1310 nm pump since the former is absorbed both in the QWs and its barriers whereas the latter is only absorbed in the available states within the QWs; thus 980 nm laser pumping was used for further characterization.

\iftrue
\begin{figure}[!t]
\centering
\includegraphics[width=3in]{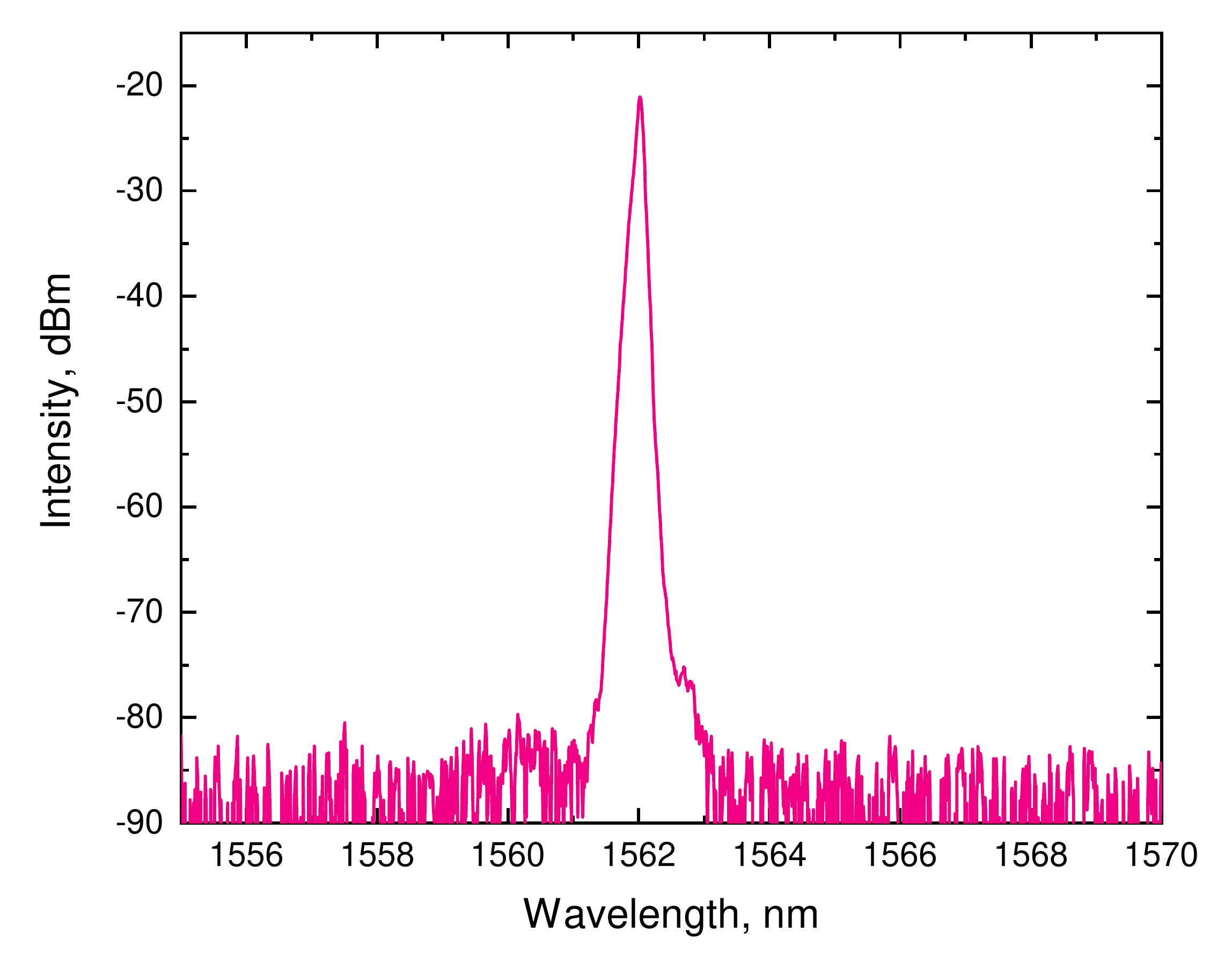}
\caption{Optical spectrum from the MEMS VCSEL using 976 nm optical pumping.}
\label{fig_spec}
\end{figure}
\fi

\subsection{DC Characteristics}
\begin{figure}[!t]
\centering
\includegraphics[width=3.2in]{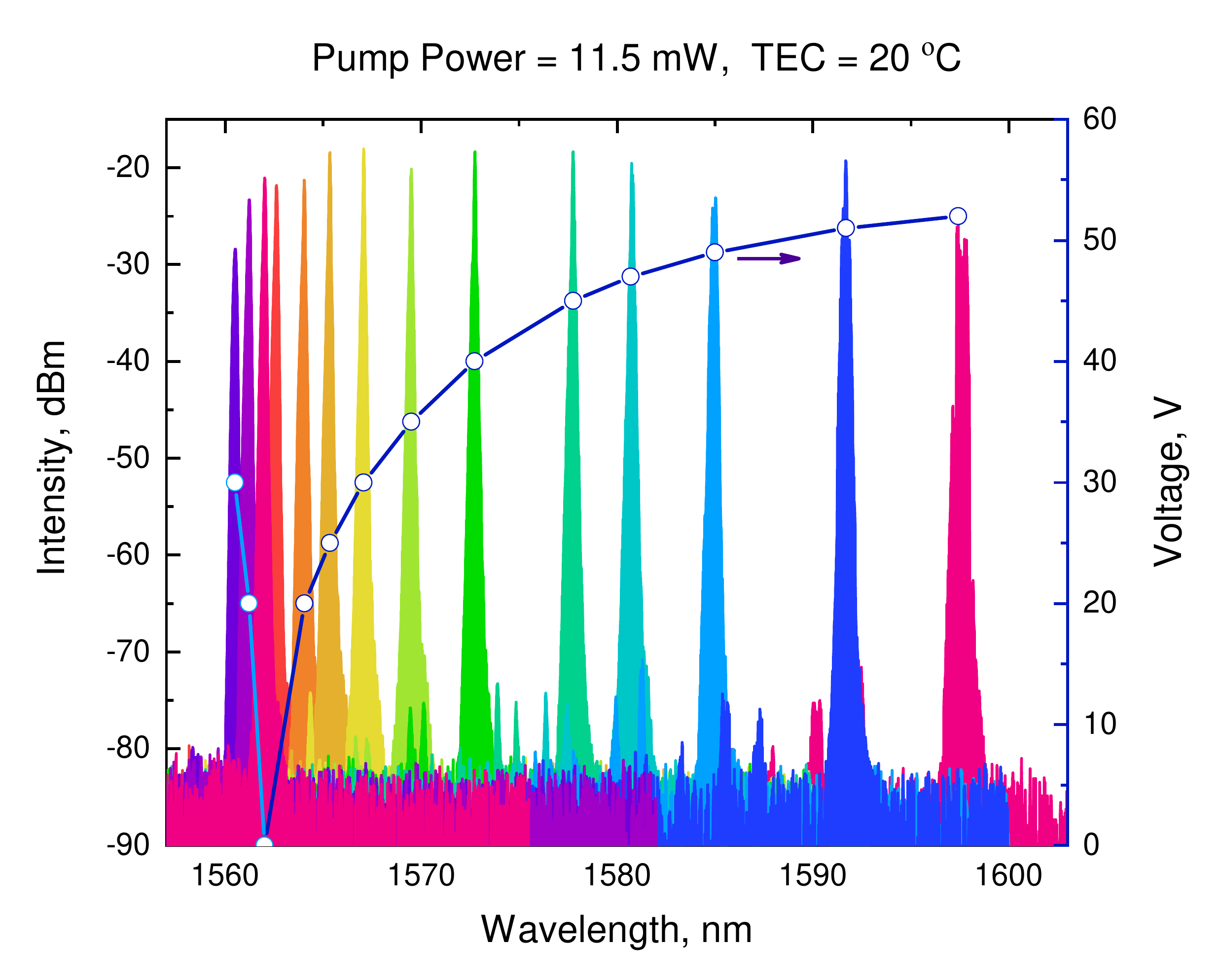}
\caption{Optical spectra showing wavelength tuning achieved for different actuation voltages. A 976 nm laser diode was used for pumping. The electrodes used for the right part of tuning were the MEMS silicon layer and the silicon substrate. The electrodes for the left part of actuation were doped InP and MEMS silicon layer. The MEMS silicon layer was grounded in both configurations.}
\label{fig_tunew}
\end{figure}

%It is possible to actuate the MEMS  by applying voltage between the centre electrode (HCG MEMS silicon layer) and either top (doped InP) or bottom (silicon substrate) electrode. Figure \ref{fig_tunew} shows the output spectrum when the MEMS is actuated. By application of a voltage between the centre and bottom electrodes, the peak emission red shifts.  Some degree of bidirectional tuning is also demonstrated as the emission wavelength blue shifts when voltage is applied between the top and centre electrodes. A very limited tuning was achieved using the top actuation either because of increased losses from heating or being at the edge of gain region. Overall, the device demonstrated a wide tuning range of approx. 40 nm.  This is, however, no way  a limit to the maximum tuning range. The active layer can be optimized to have a wider gain bandwidth. At the same time, heat extraction needs to be optimized to lower the losses to further extent the tuning range.  
The MEMS can be actuated by applying voltage between the centre electrode (HCG MEMS silicon layer) and either top (doped InP) or bottom (silicon substrate) electrodes. Fig.~\ref{fig_tunew} shows the output spectrum when the MEMS is actuated. By application of a voltage between the centre and bottom electrodes, the peak emission red shifts.  Some degree of bidirectional tuning is also demonstrated since the emission wavelength blue shifts when voltage is applied between the top and centre electrodes. A very limited tuning range was achieved using top actuation either because of increased losses from heating or gain limitations. Overall, the device demonstrated a wide tuning range of approximately 40~nm, this is, however, not a fundamental limit to the maximum tuning range. The active layer can be optimized to have a wider gain bandwidth, and at the same time, heat extraction needs to be optimized to reduce the losses to further expand the tuning range.

\subsection{AC Characteristics}
\begin{figure}[!t]
\centering
\includegraphics[width=3.2in]{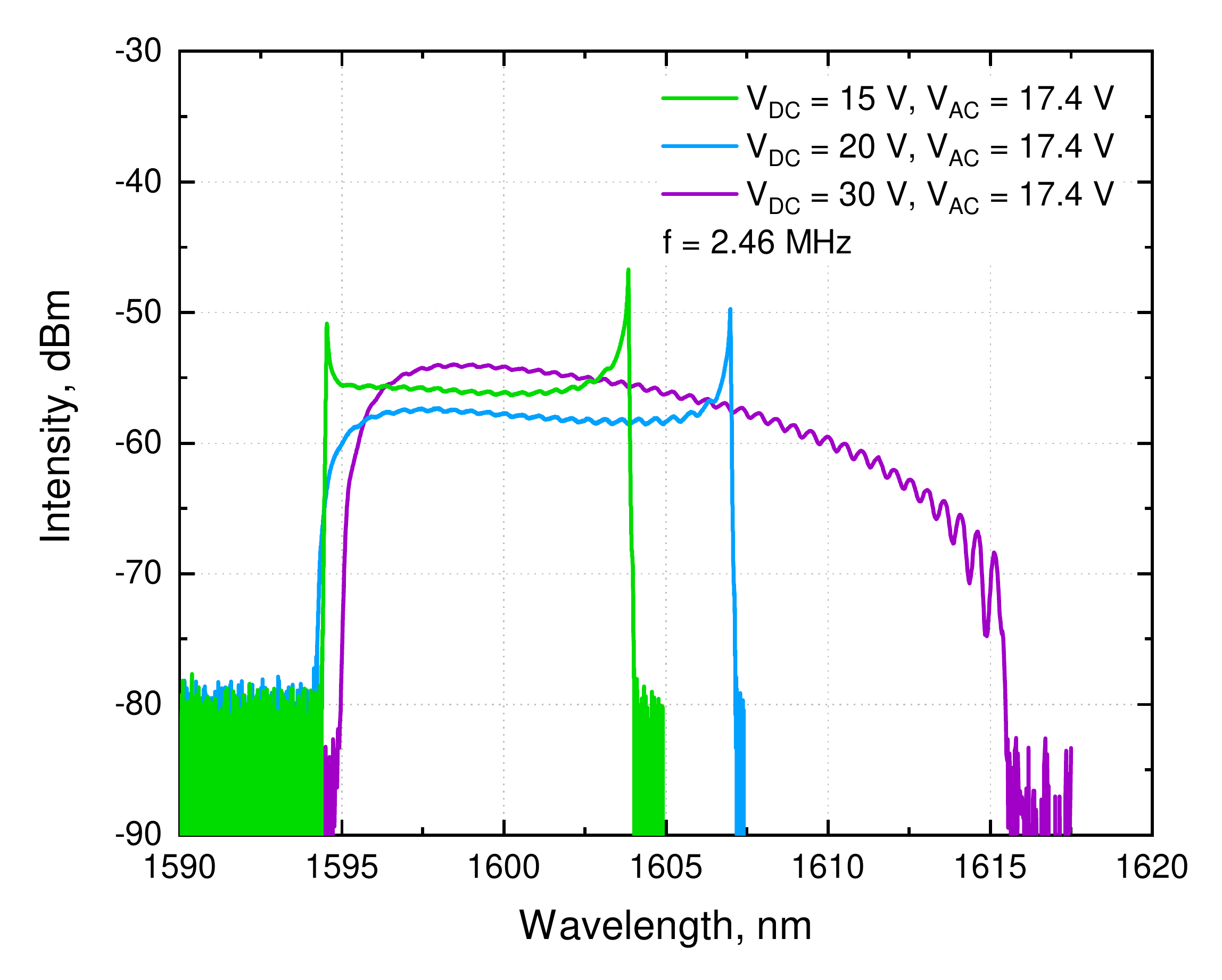}
\caption{Optical spectrum obtained for AC actuation. A voltage waveform  
$V(t)=V_{DC}+V_{AC}\sin(\omega t)$ where $\omega =2\pi f$ and $f=2.46$~MHz was applied between MEMS silicon "device" layer and the silicon substrate.}
\label{fig_tuneac}
\end{figure}

The AC dynamics of the device was studied since a continuous swept source is interesting for OCT. A voltage waveform $V(t)=V_{DC}+V_{AC}\sin(\omega t)$ where $\omega =2\pi f$, was applied between  the HCG MEMS silicon layer and the silicon substrate. The DC bias $V_{DC}$ shifts the zero point for the AC bias $V_{AC}\sin(\omega t)$ and shifts the zero position for the HCG MEMS silicon layer. Fig.~\ref{fig_tuneac} shows the AC response of the device at the resonance frequency ($f=2.46 $  MHz).  The AC actuation is studied at the resonance frequency to achieve the maximum tuning range \cite{ansbaek2013resonant}.  The use of silicon for the MEMS element helped to increase the resonance frequency. As the DC bias $V_{DC}$ is increased the centre of the tuning curve red shifts and there is a distinct increase in the bandwidth of the tuning curve. A higher $V_{DC}$ reduces the gap between the actuating electrodes, thus the electric field increases, and therefore the displacement at fixed $V_{AC}\sin(\omega t)$ increases.  AC actuation at the resonance frequency increases the tuning range beyond the achievable DC tuning range~\cite{ansbaek2013resonant}. The tuning range of around 20 nm achieved at 2.46 MHz is among the highest reported \cite{huang2008nanoelectromechanical,jayaraman2012rapidly,ansbaek2013resonant,gierl2011surface}. Different devices were used for the AC and DC actuation experiments; thus, tuning range for the MEMS VCSEL can be expected to improve in future experiments. The variation in lasing wavelength between devices came mainly from non-uniformity during the SiO$_{2}$ deposition process which defines the cavity length, thus a better uniformity control can improve the repeatability and performance in future devices.

\subsection{Thermal Effects}
\begin{figure}[!t]
\centering
\includegraphics[width=3.2in]{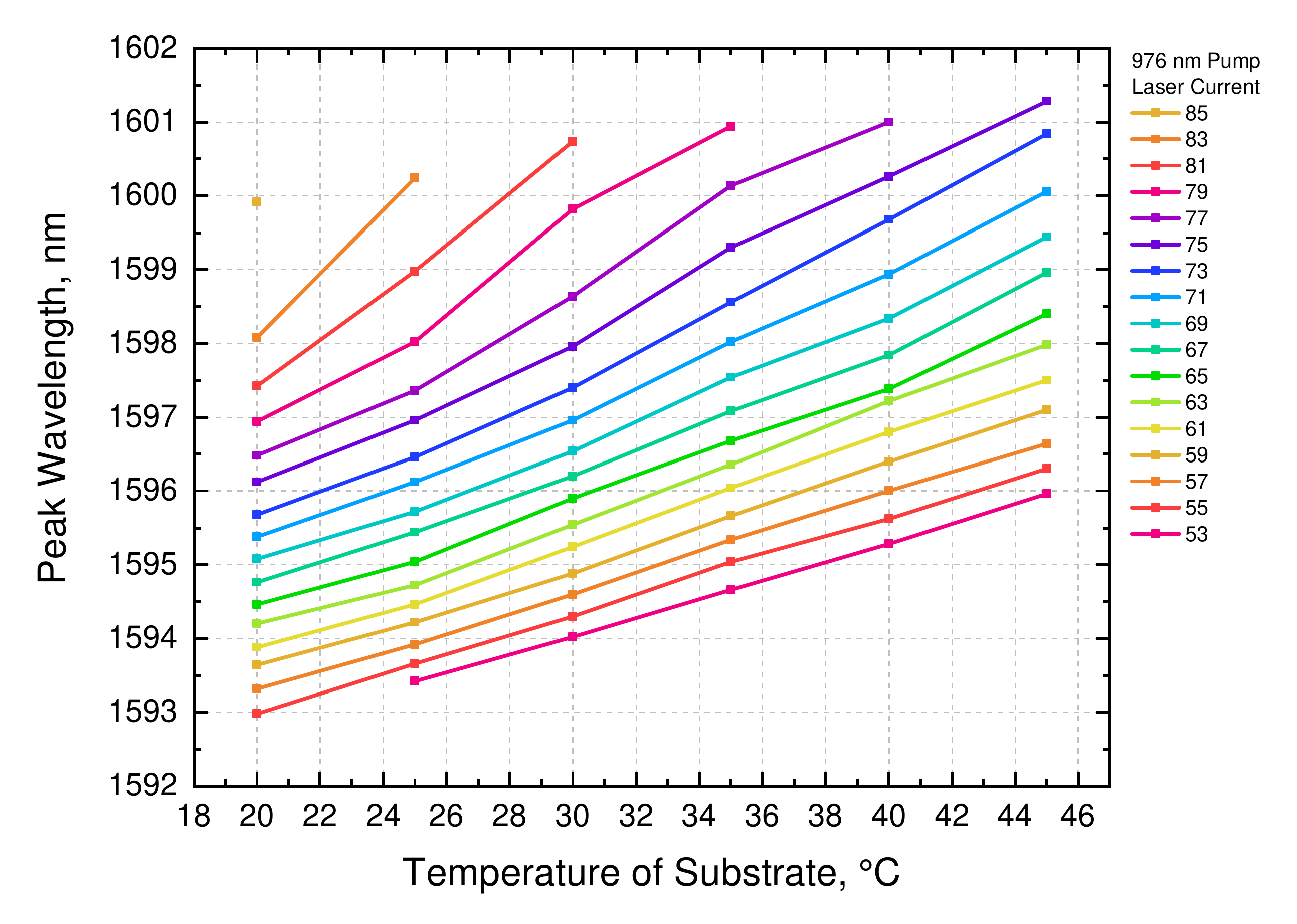}
\caption{Shift in peak lasing wavelength of the MEMS VCSEL plotted against the substrate temperature for various pump laser power. 53 mA pump laser current corresponds to the threshold for the MEMS VCSEL.}
\label{fig_heat}
\end{figure}

%Lasers are dependent on the operating temperature as it influences the gain spectrum, threshold carrier density, radiative/non-radiative recombination and overall refractive index of the system.\cite{menzel1995modelling} In addition, the device in discussion also relies on temperature for thermal lensing and, thus, lasing. The active region illuminated by the pump beam is locally heated due to the injected carriers. As the pump power is increased there is an increase in the generated carriers which increase the heat generated (non-radiative transitions). This contributes to a change in refractive index and thus shifts the lasing peak. In order to understand the effect of temperature, the emission was recorded for different pump powers at different substrate temperatures. Fig. \ref{fig_heat} shows the shift in the peak of lasing wavelength for a change in substrate temperature and pumping power. For moderate and low pumping powers, the lasing wavelength has an almost linear dependence on temperature. However, the wavelength changes (increases) much more rapidly with increase of pump power.  The slope of variation increases with pump power. Pump power, being a local phenomenon, has a more profound effect than variations with the substrate temperature. Fig. \ref{fig_heat} can be used to estimate the amount of heating on the device. As an example, for this device with a pump laser current of 69 mA, lasing at 1595 nm  has a temperature of  38 \degree C at the core, which is approx. 18 \degree C above room temperature.
Lasers are dependent on the operating temperature as it influences the gain spectrum, threshold carrier density, radiative/non-radiative recombination and overall refractive index of the system~\cite{menzel1995modelling}. In addition, the device in discussion also relies on temperature for thermal lensing and, thus, lasing. The active region illuminated by the pump beam is locally heated due to the injected carriers. As the pump power is increased carrier generation is increased which increases the heat generated (non-radiative transitions). This contributes to a change in refractive index and thus shifts the lasing peak. In order to study the effect of temperature, the emission was recorded for different pump powers at different substrate temperatures. Fig.~\ref{fig_heat} shows the shift in the peak of  the lasing wavelength for changes in substrate temperature and pumping power. For moderate and low pumping powers, the lasing wavelength has an almost linear dependence on temperature. The temperature coefficient for low pump power is close to 0.12 nm/K. However, the wavelength changes (increases) much more rapidly with increase of pump power.  The temperature coefficient increases to around 0.28 nm/K for higher pump powers. The pump power (a local phenomenon) has a more profound effect than variations with the substrate temperature. Fig.~\ref{fig_heat} can be used to estimate the amount of heating of the device due to pumping. As an example, for this device with a pump laser current of 69 mA, lasing at 1595 nm results in a temperature of  38 \degree C at the core, which is approximately 18 \degree C above room temperature.

\section{Conclusion}
%We have successfully designed and fabricated a MEMS VCSEL on a silicon substrate which encapsulates the MEMS in a cavity, enables the possibility of bi-directional tuning and is a promising robust design for future on-chip integrated devices using MEMS VCSELs. The design exploits standard silicon CMOS based processing for part of the device and all III-V processing is done after bonding to silicon. The fabricated MEMS VCSEL was characterized to have a low threshold for lasing and a  wide DC tuning range of 40 nm with some degree of bi-directional tuning. AC actuation was also explored to find the resonance frequency of MEMS and a high-frequency operation of 2.4 MHz with a tuning range of 15-20 nm was also achieved. 
We have successfully designed and fabricated a MEMS VCSEL on a silicon substrate which encapsulates the MEMS in a cavity, enables the possibility of bi-directional tuning and is a promising robust design for future on-chip integrated devices using MEMS VCSELs. The design exploits standard silicon CMOS based processing for part of the device and all III-V processing done after bonding to silicon. The fabricated MEMS VCSEL was characterized to have a low threshold for lasing and a wide DC tuning range of 40 nm with some degree of bi-directional tuning. AC actuation was also explored to find the resonance frequency of MEMS and a high-frequency operation at 2.4 MHz with a tuning range of 15-20 nm was also achieved.

\section*{Acknowledgment}

The authors would like to acknowledge the financial support from the Danish Innovation Foundation through the HERON project, the Villum Center of Excellence NATEC, and the National Basic Research Foundation center SPOC (No. DNRF123).

\ifCLASSOPTIONcaptionsoff
  \newpage
\fi



\bibliographystyle{IEEEtran}
%\bibliography{library.bib}

\begin{thebibliography}{10}
\providecommand{\url}[1]{#1}
\csname url@samestyle\endcsname
\providecommand{\newblock}{\relax}
\providecommand{\bibinfo}[2]{#2}
\providecommand{\BIBentrySTDinterwordspacing}{\spaceskip=0pt\relax}
\providecommand{\BIBentryALTinterwordstretchfactor}{4}
\providecommand{\BIBentryALTinterwordspacing}{\spaceskip=\fontdimen2\font plus
\BIBentryALTinterwordstretchfactor\fontdimen3\font minus
  \fontdimen4\font\relax}
\providecommand{\BIBforeignlanguage}[2]{{%
\expandafter\ifx\csname l@#1\endcsname\relax
\typeout{** WARNING: IEEEtran.bst: No hyphenation pattern has been}%
\typeout{** loaded for the language `#1'. Using the pattern for}%
\typeout{** the default language instead.}%
\else
\language=\csname l@#1\endcsname
\fi
#2}}
\providecommand{\BIBdecl}{\relax}
\BIBdecl

\bibitem{drexler2014optical}
W.~Drexler, M.~Liu, A.~Kumar, T.~Kamali, A.~Unterhuber, and R.~A. Leitgeb,
  ``Optical coherence tomography today: speed, contrast, and multimodality,''
  \emph{Journal of biomedical optics}, vol.~19, no.~7, pp. 071\,412--071\,412,
  2014.

\bibitem{huang1991optical}
D.~Huang, E.~A. Swanson, C.~P. Lin, J.~S. Schuman, W.~G. Stinson, W.~Chang,
  M.~R. Hee, T.~Flotte, K.~Gregory, C.~A. Puliafito \emph{et~al.}, ``Optical
  coherence tomography,'' \emph{Science (New York, NY)}, vol. 254, no. 5035, p.
  1178, 1991.

\bibitem{bezerra2009intracoronary}
H.~G. Bezerra, M.~A. Costa, G.~Guagliumi, A.~M. Rollins, and D.~I. Simon,
  ``Intracoronary optical coherence tomography: a comprehensive review:
  clinical and research applications,'' \emph{JACC: Cardiovascular
  Interventions}, vol.~2, no.~11, pp. 1035--1046, 2009.

\bibitem{suter2011intravascular}
M.~J. Suter, S.~K. Nadkarni, G.~Weisz, A.~Tanaka, F.~A. Jaffer, B.~E. Bouma,
  and G.~J. Tearney, ``Intravascular optical imaging technology for
  investigating the coronary artery,'' \emph{JACC: Cardiovascular Imaging},
  vol.~4, no.~9, pp. 1022--1039, 2011.

\bibitem{welzel2001optical}
J.~Welzel, ``Optical coherence tomography in dermatology: a review,''
  \emph{Skin Research and Technology}, vol.~7, no.~1, pp. 1--9, 2001.

\bibitem{tearney1997optical}
G.~Tearney, M.~Brezinski, J.~Southern, B.~Bouma, S.~Boppart, and J.~Fujimoto,
  ``Optical biopsy in human urologic tissue using optical coherence
  tomography,'' \emph{The Journal of urology}, vol. 157, no.~5, pp. 1915--1919,
  1997.

\bibitem{fujimoto2003optical}
J.~G. Fujimoto, ``Optical coherence tomography for ultrahigh resolution in vivo
  imaging,'' \emph{Nature biotechnology}, vol.~21, no.~11, pp. 1361--1367,
  2003.

\bibitem{liu2004rapid}
X.~Liu, M.~J. Cobb, Y.~Chen, M.~B. Kimmey, and X.~Li, ``Rapid-scanning
  forward-imaging miniature endoscope for real-time optical coherence
  tomography,'' \emph{Optics letters}, vol.~29, no.~15, pp. 1763--1765, 2004.

\bibitem{boppart1997forward}
S.~Boppart, B.~E. Bouma, C.~Pitris, G.~J. Tearney, J.~G. Fujimoto, and
  M.~Brezinski, ``Forward-imaging instruments for optical coherence
  tomography,'' \emph{Optics letters}, vol.~22, no.~21, pp. 1618--1620, 1997.

\bibitem{konig2009clinical}
K.~K{\"o}nig, M.~Speicher, R.~B{\"u}ckle, J.~Reckfort, G.~McKenzie, J.~Welzel,
  M.~J. Koehler, P.~Elsner, and M.~Kaatz, ``Clinical optical coherence
  tomography combined with multiphoton tomography of patients with skin
  diseases,'' \emph{Journal of Biophotonics}, vol.~2, no. 6-7, pp. 389--397,
  2009.

\bibitem{vinegoni2004nonlinear}
C.~Vinegoni, J.~S. Bredfeldt, D.~L. Marks, and S.~A. Boppart, ``Nonlinear
  optical contrast enhancement for optical coherence tomography,'' \emph{Optics
  express}, vol.~12, no.~2, pp. 331--341, 2004.

\bibitem{zhang2011multimodal}
E.~Z. Zhang, B.~Povazay, J.~Laufer, A.~Alex, B.~Hofer, B.~Pedley,
  C.~Glittenberg, B.~Treeby, B.~Cox, P.~Beard \emph{et~al.}, ``Multimodal
  photoacoustic and optical coherence tomography scanner using an all optical
  detection scheme for 3d morphological skin imaging,'' \emph{Biomedical optics
  express}, vol.~2, no.~8, pp. 2202--2215, 2011.

\bibitem{ansbaek2013resonant}
T.~Ansb{\ae}k, I.-S. Chung, E.~S. Semenova, O.~Hansen, and K.~Yvind, ``Resonant
  mems tunable vcsel,'' \emph{IEEE Journal of Selected Topics in Quantum
  Electronics}, vol.~19, no.~4, pp. 1\,702\,306--1\,702\,306, 2013.

\bibitem{jayaraman2012high}
V.~Jayaraman, G.~Cole, M.~Robertson, A.~Uddin, and A.~Cable, ``High-sweep-rate
  1310 nm mems-vcsel with 150 nm continuous tuning range,'' \emph{Electronics
  letters}, vol.~48, no.~14, pp. 867--869, 2012.

\bibitem{ansbaek2012vertical}
T.~Ansb{\ae}k, K.~Yvind, I.-S. Chung, and D.~Larsson, ``Vertical-cavity
  surface-emitting lasers for medical diagnosis,'' Ph.D. dissertation,
  Technical University of DenmarkDanmarks Tekniske Universitet, Department of
  Photonics EngineeringInstitut for Fotonik, NanophotonicsNanofotonik, 2012.

\bibitem{ansbaek20131060}
T.~Ansb{\ae}k, I.-S. Chung, E.~S. Semenova, and K.~Yvind, ``1060-nm tunable
  monolithic high index contrast subwavelength grating vcsel,'' \emph{IEEE
  Photonics Technology Letters}, vol.~25, no.~4, pp. 365--367, 2013.

\bibitem{jayaraman2012rapidly}
V.~Jayaraman, G.~Cole, M.~Robertson, C.~Burgner, D.~John, A.~Uddin, and
  A.~Cable, ``Rapidly swept, ultra-widely-tunable 1060 nm mems-vcsels,''
  \emph{Electronics letters}, vol.~48, no.~21, pp. 1331--1333, 2012.

\bibitem{rao2013long}
Y.~Rao, W.~Yang, C.~Chase, M.~C. Huang, D.~P. Worland, S.~Khaleghi, M.~R.
  Chitgarha, M.~Ziyadi, A.~E. Willner, and C.~J. Chang-Hasnain,
  ``Long-wavelength vcsel using high-contrast grating,'' \emph{IEEE Journal of
  Selected Topics in Quantum Electronics}, vol.~19, no.~4, pp.
  1\,701\,311--1\,701\,311, 2013.

\bibitem{chang2010high}
C.~J. Chang-Hasnain, ``High-contrast gratings as a new platform for integrated
  optoelectronics,'' \emph{Semiconductor Science and Technology}, vol.~26,
  no.~1, p. 014043, 2010.

\bibitem{lavrinenko2014numerical}
A.~V. Lavrinenko, J.~L{\ae}gsgaard, N.~Gregersen, F.~Schmidt, and
  T.~S{\o}ndergaard, \emph{Numerical methods in photonics}.\hskip 1em plus
  0.5em minus 0.4em\relax CRC Press, 2014, vol.~1.

\bibitem{bienstman2001rigorous}
P.~Bienstman, ``Rigorous and efficient modelling of wavelenght scale photonic
  components,'' Ph.D. dissertation, Ghent University, 2001.

\bibitem{bienstman2001optical}
P.~Bienstman and R.~Baets, ``Optical modelling of photonic crystals and vcsels
  using eigenmode expansion and perfectly matched layers,'' \emph{Optical and
  Quantum Electronics}, vol.~33, no. 4-5, pp. 327--341, 2001.

\bibitem{cook2019resonant}
K.~T. Cook, P.~Qiao, J.~Qi, L.~A. Coldren, and C.~J. Chang-Hasnain,
  ``Resonant-antiresonant coupled cavity vcsels,'' \emph{Optics Express},
  vol.~27, no.~3, pp. 1798--1807, 2019.

\bibitem{sahoo2018low}
H.~K. Sahoo, L.~Ottaviano, Y.~Zheng, O.~Hansen, and K.~Yvind, ``Low temperature
  bonding of heterogeneous materials using al2o3 as an intermediate layer,''
  \emph{Journal of Vacuum Science \& Technology B, Nanotechnology and
  Microelectronics: Materials, Processing, Measurement, and Phenomena},
  vol.~36, no.~1, p. 011202, 2018.

\bibitem{chung2008subwavelength}
I.-S. Chung, J.~Mork, P.~Gilet, and A.~Chelnokov, ``Subwavelength
  grating-mirror vcsel with a thin oxide gap,'' \emph{IEEE Photonics Technology
  Letters}, vol.~20, no.~2, pp. 105--107, 2008.

\bibitem{huang2008nanoelectromechanical}
M.~C. Huang, Y.~Zhou, and C.~J. Chang-Hasnain, ``A nanoelectromechanical
  tunable laser,'' \emph{Nature Photonics}, vol.~2, no.~3, p. 180, 2008.

\bibitem{gierl2011surface}
C.~Gierl, T.~Gruendl, P.~Debernardi, K.~Zogal, C.~Grasse, H.~Davani,
  G.~B{\"o}hm, S.~Jatta, F.~K{\"u}ppers, P.~Mei{\ss}ner \emph{et~al.},
  ``Surface micromachined tunable 1.55 $\mu$m-vcsel with 102 nm continuous
  single-mode tuning,'' \emph{Optics Express}, vol.~19, no.~18, pp.
  17\,336--17\,343, 2011.

\bibitem{menzel1995modelling}
U.~Menzel, A.~Barwolff, P.~Enders, D.~Ackermann, R.~Puchert, and M.~Voss,
  ``Modelling the temperature dependence of threshold current, external
  differential efficiency and lasing wavelength in qw laser diodes,''
  \emph{Semiconductor science and technology}, vol.~10, no.~10, p. 1382, 1995.

\end{thebibliography}

%\newpage

\begin{IEEEbiographynophoto}{Hitesh Kumar Sahoo}  is currently working as a post-doc researcher with the High-Speed Optical Communication group at the Technical University of Denmark (DTU) Fotonik. He received B.E. (Hons.) in Electronics and Instrumentation from Birla Institute of Technology and Science - Goa, India in 2011, M.Sc. in Nanotechnology from the University of Pennsylvania, USA, in 2013 and PhD in the fabrication of wavelength tunable MEMS VCSELs from DTU in 2018. His research interests include MEMS, heterogeneous material integration, nano-fabrication and integrated optics. 
\end{IEEEbiographynophoto}

\begin{IEEEbiographynophoto}{Thor Ansb{\ae}k}
 was born in Copenhagen, Denmark, in 1984. He received the M.Sc. Eng. degree in 2008 and the Ph.D. degree in 2012 both at The Technical University of Denmark, Kongens Lyngby, Denmark. His research interests include semiconductor fabrication, microelectromechanical systems, and vertical-cavity surface-emitting lasers for medical diagnosis. Currently CEO of OCTLIGHT ApS, a company developing OCT Swept Sources for Optical Coherence Tomography.
\end{IEEEbiographynophoto}

%\newpage

\begin{IEEEbiographynophoto}{Luisa Ottaviano}
 holds a PhD within physics of semiconductor by the University of Catania and is currently working as development engineer at Alight technologies where she is mainly responsible for the process development, characterisation and failure mode analysis of VCSEL devices for telecommunication. From 2008 to 2017 she has worked at the Department of Photonics Engineer at the Technical University of Denmark (DTU) first as Post Doc, then as academic technician where she has been involved in several projects involving cleanroom process development, optimization and fabrication of devices based on III-V compound semiconductors. She is author or co-author of 50 peer reviewed articles and conference contributions.
\end{IEEEbiographynophoto}

\begin{IEEEbiographynophoto}{Elizaveta Semenova}
 received her B.Sc and M.Sc degrees from St.-Petersburg State Technical University, Russia in 1999 and 2001, respectively, and she received her PhD degree in semiconductor physics from A.F. Ioffe Institute, Russia in 2005. Currently she is holding the position of a Senior Researcher at the Technical University of Denmark at DTU Fotonik. Her research interests are focused on the development and the optimization of the epitaxial growth process of new III-V semiconductor materials for optoelectronic applications, including III-V on Si. She is a co-author of more than 76 peer-reviewed scientific publications.
\end{IEEEbiographynophoto}

\begin{IEEEbiographynophoto}{Fyodor Zubov}
 was born in Leningrad (now Saint Petersburg, Russia) in 1984. He received an MSc degree in technical physics from St. Petersburg State Polytechnical University, St. Petersburg, Russia, in 2008. Since 2008 till 2010 he was with Phystex, Netherlands, as a research fellow, working on radiation source for extreme ultraviolet lithography. From 2010 to date he is a research fellow at Nanophotonics lab. at St. Petersburg Academic University. In 2013 he received the Ph.D. degree from St. Petersburg Academic University. His current research interests include diode lasers with asymmetric barrier layers, semiconductor lasers based on microresonators and terahertz quantum cascade lasers.
\end{IEEEbiographynophoto}

\begin{IEEEbiographynophoto}{Ole Hansen}
 is Professor at DTU Nanotech, the Technical University of Denmark, where he is heading the Silicon Microtechnology group, with activities within lithography based micro- and nano-technology. He received his MSc degree within micro-technology from the Technical University of Denmark in 1977, and has since then worked with micro- and nano-technology and applications of the technology within electronics, metrology, sensing, catalysis and energy harvesting. Current research interests include sustainable energy, photo-catalysis and tools for characterizing catalytic processes. From 2005-2016 he was part of the Danish National Research Foundation Center CINF, Center for Individual Nanoparticle Functionality, and since 2016 he has been part of V-Sustain, the Villum Center for the Science of Sustainable Fuels  and Chemicals. 

\end{IEEEbiographynophoto}

\begin{IEEEbiographynophoto}{Kresten Yvind}
 received the M.Sc.E. and Ph.D. degree in 1999 and 2003 from the Research Center for Communication, Optics and Materials (COM) at the Technical University of Denmark. The center was renamed DTU Fotonik in 2008 and he is currently associate professor there. His work is centered on III-V and silicon optoelectronic devices and involves design of epitaxial structures to growth, processing and high-speed characterization. Examples of work has been passive high contrast waveguide devices and high-speed functional waveguide elements i.e. mode-locked lasers, electro-absorption modulators and semiconductor optical amplifiers for applications in optical communication systems or microwave photonics and MEMS VCSELs for optical coherence tomography. Membrane based devices (on silicon) have been a focus the last decade leading to efficient nonlinear integrated photonics, various (active) photonics crystal devices for optical interconnects and MEMS VCSELs.
\end{IEEEbiographynophoto}

\vfill

\enlargethispage{-5in}
\newpage

\end{document}